\documentstyle[epsfig,referee]{mn}

\newif\ifAMStwofonts
\ifoldfss
  \ifCUPmtlplainloaded \else
    \NewTextAlphabet{textbfit} {cmbxti10} {}
    \NewTextAlphabet{textbfss} {cmssbx10} {}
    \NewMathAlphabet{mathbfit} {cmbxti10} {} % for math mode
    \NewMathAlphabet{mathbfss} {cmssbx10} {} %  "   "    "
  \fi
  \ifAMStwofonts
    \ifCUPmtlplainloaded \else
      \NewSymbolFont{upmath} {eurm10}
      \NewSymbolFont{AMSa} {msam10}
      \NewMathSymbol{\upi}     {0}{upmath}{19}
      \NewMathSymbol{\umu}     {0}{upmath}{16}
      \NewMathSymbol{\upartial}{0}{upmath}{40}
      \NewMathSymbol{\leqslant}{3}{AMSa}{36}
      \NewMathSymbol{\geqslant}{3}{AMSa}{3E}

       \let\le=\leqslant
       \let\ge=\geqslant
    \fi
  \fi
\fi % End of OFSS

\ifnfssone
  \newmathalphabet{\mathit}
  \addtoversion{normal}{\mathit}{cmr}{m}{it}
  \addtoversion{bold}{\mathit}{cmr}{bx}{it}
  \newmathalphabet{\mathbfit} % math mode version of \textbfit{..}
  \addtoversion{normal}{\mathbfit}{cmr}{bx}{it}
  \addtoversion{bold}{\mathbfit}{cmr}{bx}{it}
  \newmathalphabet{\mathbfss} % math mode version of \textbfss{..}
  \addtoversion{normal}{\mathbfss}{cmss}{bx}{n}
  \addtoversion{bold}{\mathbfss}{cmss}{bx}{n}
  \ifAMStwofonts
    \ifCUPmtlplainloaded \else
      \UseAMStwoboldmath
      \makeatletter
      \new@mathgroup\upmath@group
      \define@mathgroup\mv@normal\upmath@group{eur}{m}{n}
      \define@mathgroup\mv@bold\upmath@group{eur}{b}{n}
      \edef\UPM{\hexnumber\upmath@group}
      \new@mathgroup\amsa@group
      \define@mathgroup\mv@normal\amsa@group{msa}{m}{n}
      \define@mathgroup\mv@bold\amsa@group{msa}{m}{n}
      \edef\AMSa{\hexnumber\amsa@group}
      \makeatother
      \mathchardef\upi="0\UPM19
      \mathchardef\umu="0\UPM16
      \mathchardef\upartial="0\UPM40
      \mathchardef\leqslant="3\AMSa36
      \mathchardef\geqslant="3\AMSa3E

       \let\le=\leqslant
       \let\ge=\geqslant
    \fi
  \fi
\fi % End of NFSS release 1

\ifnfsstwo
  \DeclareMathAlphabet{\mathbfit}{OT1}{cmr}{bx}{it}
  \SetMathAlphabet\mathbfit{bold}{OT1}{cmr}{bx}{it}
  \DeclareMathAlphabet{\mathbfss}{OT1}{cmss}{bx}{n}
  \SetMathAlphabet\mathbfss{bold}{OT1}{cmss}{bx}{n}
  \ifAMStwofonts
    \ifCUPmtlplainloaded \else
      \DeclareSymbolFont{UPM}{U}{eur}{m}{n}
      \SetSymbolFont{UPM}{bold}{U}{eur}{b}{n}
      \DeclareSymbolFont{AMSa}{U}{msa}{m}{n}
      \DeclareMathSymbol{\upi}{0}{UPM}{"19}
      \DeclareMathSymbol{\umu}{0}{UPM}{"16}
      \DeclareMathSymbol{\upartial}{0}{UPM}{"40}
      \DeclareMathSymbol{\leqslant}{3}{AMSa}{"36}
      \DeclareMathSymbol{\geqslant}{3}{AMSa}{"3E}

       \let\le=\leqslant
       \let\ge=\geqslant
    \fi
  \fi
\fi % End of NFSS release 2

\ifCUPmtlplainloaded \else
  \ifAMStwofonts \else % If no AMS fonts
    \def\upi{\pi}
    \def\umu{\mu}
    \def\upartial{\partial}
  \fi
\fi

\title{The new model of a tidally disrupted star: further development and relativistic
calculations}                       

\author[P. B. Ivanov et al.]
  { P. B. Ivanov, $^{1,4}$, M. A. Chernyakova,$^{2,4,5}$ 
  I. D. Novikov, $^{3,4,6,7}$\\
  $^1$ School of Mathematical Sciences, Queen Mary, University of
London, Mile End Road, London E1 4NS, United Kingdom \\
  $^2$ Integral Science Data Center, Chemin d'Ecogia 16, CH-1290 Versoix,
Switzerland\\
  $^3$ Theoretical Astrophysics Center, Juliane Maries Vej 30,
 DK-2100 Copenhagen, Denmark\\ 
  $^4$  Astro Space Center of PN Lebedev Physical Institute, 84/32 
Profsoyuznaya Street, Moscow, 117810, Russia \\
 $^5$ Geneva Observatory, CH-1290, Sauverny, Switzerland \\
 $^6$ University Observatory, Juliane Maries Vej 30,
 DK-2100 Copenhagen, Denmark \\ 
 $^7$ NORDITA, Blegdamsvej 17, DK-2100, Copenhagen, Denmark}

\pagerange{\pageref{firstpage}--\pageref{lastpage}}
\pubyear{1994}

\begin{document}

\maketitle

\label{firstpage}

\begin{abstract}

\end{abstract}
In this paper we develop the new semi-analytical model of 
a tidally perturbed or tidally disrupted star proposed recently
by two of us. This model is effectively a one dimensional Lagrangian
model and it can be evolved numerically much faster than
the conventional $3D$ models.
A self-consistent derivation of dynamical equations of
the model is performed and several important theorems about 
the dynamics of the model are proved without any particular assumption
about the equation of state of the stellar gas. The dynamical equations
are solved numerically for the case of an $n=1.5$ polytropic star
evolving in the relativistic 
tidal field of a $10^{7}M_{\odot}$ Kerr black hole. Some results of these
calculations are compared with the results of calculations based on
finite-difference $3D$ simulations. The comparison shows a very
good agreement between both approaches to the problem. 
Then we show that the strength of the tidal encounter 
depends significantly on the relative orientation of the orbital
angular momentum of the star and the spin of the black hole.     
\begin{keywords}
black hole physics - galaxies: nuclei
\end{keywords}

\section{Introduction}

This paper continues studying a new dynamical model of a star
evolving under the influence of a tidal field. This  
semi-analytical model has been proposed 
by two of us in a recent paper (Ivanov $\&$ Novikov, 2001, hereafter IN).
It allows researchers to calculate the outcome of a strong tidal encounter of a
star with a source of a tidal field  much faster than the standard 
three-dimensional approach, and evolves the star under the influence of 
a tidal field during a much longer time.  On the other hand, testing of
the model for the case of a Newtonian tidal field of a point mass has shown
that the new model gives very good quantitative agreement with the results of 
3D simulations. Therefore, the new model could be used in a situation where
the formulation of a problem demands many different computations of
the tidal encounter events with different parameters, or for calculating
the stellar evolution in a complicated 
tidal field, and where
the present day 3D simulations cannot be used due to problems with 
computational time or other problems
\footnote{
See IN for an overview of works on tidal encounters and astrophysical 
applications.}. 
A natural example of such a problem is an attempt of surveying
the parameter space of the 
problem of tidal disruption of stars by a Kerr black hole.

The main feature of the new model consists of an assumption about the motion
of the stellar gas perturbed by the action of a tidal field. 
Namely, it is assumed that different mass shells
of the star always
keep the shape of ellipsoids during the evolution of the star in the tidal field.
\footnote{
Therefore, the model is a direct generalisation of the so-called affine model
of the star (Carter $\&$ Luminet, 1982, 1983, 1985,
Lattimer $\&$ Schramm, 1976) which has been intensively used for investigation 
of the tidal
encounters before the development of modern 3D computational methods. 
However, contrary to the affine model, the different elliptical mass shells evolve 
differently, with different parameters and orientations.}
This assumption
allows us to reduce the complicated non-linear three dimensional dynamics of
the stellar gas to an effectively one-dimensional Lagrangian numerical scheme.
The dynamical equations 
of our model are derived from the so-called virial relations written for each
mass shell (see the next Section), and form a set of non-linear one-dimensional
partial differential equations of hyperbolic type coupled with the tidal field. 
They depend on time and a Lagrangian variable which could be represented
by the mass enclosed within a particular shell or the radius of the shell in
the unperturbed spherical state of a star.

There is one fundamental drawback in the variant of the model studied by IN. 
Namely, the 'thermal terms' (i.e. the terms determined by the gas energy and
pressure) in the virial relations were treated by IN in an approximate manner.
This led to unphysical behaviour of the model, the mass shells corresponding
to different Lagrangian radii were allowed to intersect each other during the
evolution of the star. Therefore, the variant of IN was not suitable for the study
of the density, pressure and velocity distribution within the star. However, the
distribution of these quantities represents a significant interest in the problems
connected with the problem of tidal disruption. For example, one would need to
know these quantities for a study of the subsequent evolution of the gas liberated
from the star after a fly-by of a black hole and the formation of an accretion disc
(or torus). The main purpose of this paper is to resolve this difficulty of the
model. In this paper, we calculate the 'thermal terms' exactly and derive 
the dynamical equations of our model in a self-consistent way (see the next Section).
Then, we apply our advanced variant of the model to the problem of a fly-by of
a $n=1.5$ polytropic star around a Kerr black hole. We test the model against 
three dimensional calculations made by Diener et al (Diener et al, 1997) for
the same problem and the same parameters. We find very good agreement between
our calculations and the calculations based on the 3D approach. 
   
Our paper is organised as follows. We derive the dynamical equations of our 
model in the next Section. In Section 3 we discuss the results of 
numerical simulations. Discussion and conclusions are presented in Section
4. The paper is written in a self-consistent way, and all important relations
are derived in the text without referring to IN. 

Following IN we use an unusual summation convention: summation is performed  
over all indices appearing in our expressions more than once but summation is not
performed if indices are enclosed in brackets. Bold letters represent matrices
in abstract form. All indices can be raised or lowered with help of the Kronecker
delta symbol, but we distinguish between the upper and lower indices in order to
enumerate the rows and columns of matrices, respectively.

\section{Derivation and analysis of the dynamical equations}

We derive and analyse  dynamical equations of our model
in a manner similar to that was described in our previous paper (IN).
However, as we mention in the Introduction, 
we do not use simplifying assumptions concerning ``thermal
terms'' in our equations (for the exact definition of the thermal
terms, see equations (15-19) below). Also, we prove several important theorems
about the dynamics of our model without any particular assumptions about the
equation of state of the stellar gas. 

At first, we would like to introduce coordinate systems and
several useful kinematical quantities. We use two different coordinate systems:
a) Cartesian coordinates $x^{i}$ associated with a locally inertial frame centered at
the star's geometrical center (we call those below ``the Eulerian coordinates'' of
the gas element); b) Cartesian coordinates $x^{i}_{0}$ of the gas elements
in an unperturbed spherical state  of the star (say, before the star is deformed by the
tidal field). By definition, these coordinates are not changing during the evolution of
any particular gas element, and therefore we call them below  ``the
Lagrangian coordinates''. As we have mentioned in the Introduction,  
we assume that the star 
consists of elliptical shells, and these shells are not deformed during the
evolution of the star. This assumption allows us to write the law of
transformation between the Lagrangian and Eulerian coordinates in the form:
$$x^{i}=T^{i}_{j}(t, r_{0})e_{0}^{j}, \eqno 1$$
where $r_{0}=\sqrt{x_{0i}x^{i}_{0}}$ is the Lagrangian radius of a particular
shell, and 
$e_{0}^{i}=x^{i}_{0}/r_{0}$ are direction cosines in the Lagrangian space 
($e_{0i}e^{i}_{0}=1$). We represent
the position matrix ${\bf T}$ and its inverse
${\bf S}$ as a product of
two rotational matrices ${\bf A}$ and ${\bf E}$, and a diagonal 
matrix ${\bf B}$:
$$T^{i}_{j}=A^{i}_{l}B^{l}_{m}E^{m}_{j}=a_{l}A^{i}_{l}E^{l}_{j}, 
\quad S^{i}_{j}=a_{l}^{-1}A^{j}_{l}E^{l}_{i}, \eqno 2$$
where $B^{l}_{m}=a_{(l)}\delta^{(l)}_{m}$, and $a_{l}$ are the principal
axes of the elliptical shell.  
The Jacobian 
 $D=|{\partial x^{i} \over \partial x_{0}^{j}}|$ of the
mapping between the Lagrangian and Eulerian spaces can be written as
$$D(x_{0}^{i})={ge_{0l}e_{n0}R^{ln} \over
r_{0}^{2}}, \eqno 3$$
where
$$g=|{\bf T}|=a_{1}a_{2}a_{3}, \eqno 4$$
is the determinant of the matrix ${\bf T}$, and
the symmetric matrix $R^{ln}$ determines a local shear and a change of 
volume of the neighbouring shells:
$$R^{ln}={1\over 2}(S^{l}_{m}(T^{m}_{n})\prime + S^{n}_{m}(T^{m}_{l})\prime), \eqno 5$$ 
with the primes standing for differentiation with respect to $r_{0}$
\footnote{Note a useful formula for the averaged value of the
Jacobian $D$ (IN): $\bar D={1\over 4\pi}\int d\Omega D={dg\over dr_{0}^{3}}$.
Here integration is performed over a unit sphere in the Lagrangian space.}.
The matrix $\bf R$ can be represented in terms of its eigenvalues $f_{m}$ and
the rotational matrix $\bf O$:
$$R^{ln}=f_{m}O^{l}_{m}O^{n}_{m}. \eqno 6$$
As it follows from the law of mass conservation, the evolution of the gas density
$\rho$ is determined by the evolution of $D$:
$$\rho(t,x^{i})={\rho_{0}(r_{0})\over D}, \eqno 7$$
where $\rho_{0}(r_{0})$ is the gas density in the unperturbed spherical state of
the star. If one of the eigenvalues $f_{m}$ goes to zero, 
the density $\rho(t,x^{i})$ goes to infinity at a certain value of 
$x^{i}_{0}$. This physically corresponds to the intersection of two neighbouring shells.
However, we assume that pressure forces can always prevent the shells from intersecting,
and therefore the eigenvalues $f_{m}$ are always positive. 

Similar to IN, we start the derivation of the dynamical equation of our model from
the integral consequences of the exact hydrodynamical equations: the equation
of energy conservation and the so-called virial relations. We write the energy
conservation equation in adiabatic approximation, thus neglecting the energy transfer
between neighbouring shells, the entropy generation due to nuclear reactions, viscosity,
etc.,
$${d\over dt}\lbrace \int d^3x(\rho v^{2}/2 +\epsilon)+{\mathcal P}\rbrace
=-\int dS_{i}(pv^{i})+\int d^3x(\rho C_{ij}v^{i}x^{j}), \eqno 8$$
Here $v^{i}$ is the
velocity of the gas element, $v=\sqrt{v_{i}v^{i}}$, 
$p$ is the pressure and $\epsilon$ is
the energy density per unit volume. ${\mathcal P}$ stands for the potential energy
of the star. 
 The matrix $C^{i}_{j}$ represents the tidal tensor, and 
therefore it is symmetric and traceless. The virial relations have the form:   
$${d\over dt}\int d^3x(\rho x^{k}v^{i})=\int d^3x (\rho v^{k}v^{i})
+\delta^{ki}\int d^3 x p $$
$$-\int dS_{i}(x^{k}p)+{\mathcal P}^{ki}+
\int d^3x (\rho C^{i}_{j}x^{k}x^{j}), \eqno 9$$
where ${\mathcal P}^{ki}$ is the so-called potential energy tensor:
$${\mathcal P}^{ki}=-{1\over 2}\int d^3x \int d^3x_{1} \rho(x^{i})
\rho(x_{1}^{i})
{(x^{k}-x^{k}_{1})(x^{i}-x_{1}^{i})
\over |\vec x - \vec x_{1}|}. \eqno 10$$
Obviously, we have ${\mathcal P}={\mathcal P}^{kk}$. 

Now we substitute the evolution law (1) in equations (9) and calculate all terms
in these equations. Then, differentiating the result with respect to the Lagrangian
mass
coordinate 
$$M(r_{0})=4\pi \int^{r_{0}}_{0}\rho_{0}(r_{1})dr_{1}    \eqno 11$$ 
we obtain the dynamical equations of our model
(we use below the mass $M(r_{0})$ of the gas inside the shell of radius $r_{0}$ 
as a new Lagrangian coordinate instead of $r_{0}$). Analogously we obtain
a differential form of the law of energy conservation from the integral form
(8). 

We calculate explicitly ``gravitational'' parts of  equations (8) and (9) 
(the potential energy and potential energy tensor) making an additional simplifying
approximation. Namely,
we assume that the gravitational force acting
on the gas near the shell with some Lagrangian radius $r_{0}$ is equivalent
to the gravitational force of a uniform density ellipsoid with a mass equal to the part of
the star's mass within that shell. The principal
axes of that ellipsoid coincide with the principal axes of the shell,
and the density is averaged over the volume enclosed in the shell.
Under this assumption the ``averaged'' potential energy tensor 
$\bar {\mathcal P}^{ik}$ has the form
$$\bar {\mathcal P}^{ik}= -{1\over 2}\int dM GMA^{i}_{j}A^{k}_{j}
{a_{j}^{2}D_{j}\over g}, \eqno 12$$
and the ``averaged'' potential energy 
$\bar {\mathcal P}=  \bar {\mathcal P}^{kk}$ is
$$\bar {\mathcal P}= -{1\over 2}\int dM GM
{a_{j}^{2}D_{j}\over g}. \eqno 13$$
The dimensionless
quantities $D_{j}$ have been described in e. g. Chandrasekhar, 1969, p. 41.
They have the form:
$$D_{j}=g\int^{\infty}_{0}{du\over \Delta (a_{j}^{2}+u)}, \eqno 14$$
where $\Delta=\sqrt{(a_{1}^{2}+u)(a_{2}^{2}+u)(a_{3}^{2}+u)}.$

For the ``thermal'' terms $\Pi^{ik}\equiv \delta^{ik} \int d^{3}x p -\int dS_{i} x^{k}p $ 
in equation (9), we
obtain
$$\Pi^{ik}=
\delta^{ik}\int dM \bar {({p\over \rho})}-4\pi g S^{j}_{i}T^{k}_{l}{\bar P}^{jl}, 
\eqno 15$$
where 
$$\bar {({p\over \rho})}={1\over 4\pi}\int d\Omega {p\over \rho}, \eqno 16$$
and 
$${\bar P}^{jl}={1\over 4\pi}\int d\Omega pe_{0}^{j}e_{0}^{l}, \eqno 17$$
and the integration is performed over a unit sphere in 
Lagrangian space. Analogously, the energy term
$\int d^3x \epsilon$ in  equation (8) has the form:
$$\int d^3x \epsilon = \int dM \bar {({\epsilon \over \rho})}, \eqno 18$$
where $\bar {({\epsilon \over \rho})}={1\over 4\pi}\int d\Omega {\epsilon \over \rho}$,
and the surface term $\int dS_{i}(pv^{i})$ has the form:
$$\int dS_{i}(pv^{i})=4\pi g{\bar P}^{kl}S^{k}_{i}\dot T^{i}_{l}. \eqno 19$$
The calculation of other terms in equations (8), (9) is straightforward. 
Differentiating equation (8) with respect to $M$ and taking into account (13), 
(18,19), we obtain:
$${d\over dt}\lbrace  {\dot T^{i}_{n}\dot T^{i}_{n}\over 2}
+3\bar {({\epsilon\over \rho})} -{3\over 2}a^{2}_{j}D_{j}
{GM\over g}\rbrace
=-12\pi{d\over dM}\lbrace g{\bar P}^{kl}S^{k}_{i}\dot T^{i}_{l}\rbrace +C^{i}_{j}
\dot T^{i}_{l}T^{j}_{l}. \eqno 20$$
Equation (20) is a differential form of the law of energy conservation.
It is analogous to equation (22) of IN. 
Analogously, differentiating equation (9) and taking into account
(12) and (15), we obtain:
$$\ddot T^{i}_{j}=3S^{j}_{i}\bar {({p\over \rho})}-12\pi S^{j}_{k}{d\over dM}
\lbrace gS^{l}_{i}T^{k}_{n}{\bar P}^{ln}\rbrace
-{3\over 2}A^{i}_{k}a_{k}D_{k}E^{k}_{j}{GM\over g}+C^{i}_{k}T^{k}_{j}.
\eqno 21$$
Equations (21) are the dynamical equations of our model. They are
analogous to equations (23) of IN. 

Equation (20) must follow from equations (21). To prove this, we contract
both sides of equations (21) with the velocity matrix $\dot T^{i}_{n}$ over
all indices and subtract the result from equation (20). The remainder
is separated into gravitational and thermal parts. As it was shown by IN,
the gravitational part is reduced to the identity:
$${d\over dt}({a_{j}^{2}D_{j}\over g})+{D_{j}a_{j}\dot a_{j}\over g}=0.
\eqno 22$$ 
For the thermal part we have:
$${d\over dt}\bar {({\epsilon \over \rho})}+3H\bar {({p \over \rho})}+
{g\over \rho_{0}r_{0}^{2}}{\bar P}^{ln}{\dot R}_{ln}=0, \eqno 23$$
where the expansion rate is defined as
$$H={1\over 3}S^{l}_{i}\dot T^{i}_{l}={1\over 3}({\dot a_{1} \over a_{1}}  
+{\dot a_{2} \over a_{2}}+{\dot a_{3} \over a_{3}}), \eqno 24$$
and we change variables in the last term according to rule (11).
Now we are going to prove that equation (23) follows directly from
the first law of thermodynamics written in the adiabatic approximation:
$${d\over dt}({\epsilon \over \rho})+p{d\over dt} ({1 \over \rho})=0. \eqno 25$$
For that, we differentiate equations (3),(7) with respect to time to find:
$$\dot D=3HD+{g\over r_{0}^{2}}e_{0l}e_{0n}\dot R^{ln}, \eqno 26 $$
and 
$${\dot \rho \over \rho}=-{\dot D \over D} \eqno 27$$ 
Substituting (26), (27) in equation (25), and using (7), we obtain
$${d\over dt}({\epsilon \over \rho})+3H{p \over \rho}
+{g\over \rho_{0}r_{0}^{2}}Pe_{0}^{n}e_{0}^{l}{\dot R}_{ln}=0. \eqno 28$$
Integrating (28) over the solid angle $\Omega$, we obtain (23).
Therefore, equation (23) is equivalent to the zeroth moment of 
equation (25):
$${d\over dt}\bar {({\epsilon \over \rho})}+
\int {d\Omega\over 4\pi}p{d\over dt} ({1 \over \rho})=0. \eqno 29$$

Now we are going to show that the quantities 
$$ \chi_{jk}(M)=T^{i}_{k}\dot T^{i}_{j}-T^{i}_{j}\dot T^{i}_{k}, \eqno 30$$
are exact integrals of motion of the dynamical system (21). For that,
we contract the left and right hand sides of equations (21) with
$T^{i}_{k}$ and take the antisymmetric part of the result. We obtain
$$\dot \chi_{jk}(M)=12\pi g\lbrace{\bar P}^{ln}(\delta^{kn}T^{i}_{j}{d\over dM}S^{l}_{i}
+\delta^{lj}S^{k}_{\rho}{d\over dM}T^{\rho}_{n}
-\delta^{jn}T^{i}_{k}{d\over dM}S^{l}_{i}
-\delta^{kl}S^{j}_{\rho}{d\over dM}T^{\rho}_{n})\rbrace \eqno 31$$ 
The quantity in braces in (31) can be transformed to
$${\bar P}^{nj}R^{kn}-{\bar P}^{nk}R^{jn}. \eqno 32$$
To prove that (32) is equal to zero, it is sufficient to note that
both symmetric matrices $\bar {\bf P}$ and $\bf R$ can be diagonalized by the
same orthogonal transformation (defined with the help of the matrix ${\bf O}$,
see equation (6)). Therefore, the commutator (32) 
of $\bar {\bf P}$ and $\bf R$ must be equal to zero.
As was discussed by IN, the quantities $\chi_{jk}(M)$ describe the circulation
of the fluid along our elliptical shells, and therefore the conservation
law 
$$\dot \chi_{jk}(M)=0. \eqno 33$$
is a direct consequence of the conservation of circulation in our model. 

Let us discuss the law of conservation of angular momentum. In our model the
angular momentum tensor $L^{ki}$ can be expressed as
$$L^{ki}={1\over 3}\int dM (T^{k}_{j}\dot T^{i}_{j}-T^{i}_{j}\dot T^{k}_{j}). \eqno 34$$
Let the quantity $l^{ki}$ be the angular momentum tensor density per unit of mass:
$l^{ki}={d\over dM}L^{ki}$. Then, one can easily obtain the evolution law for
$l^{ki}$ from equations (21):
$$\dot l^{ki}=4\pi{d\over dM}\lbrace g{\bar P}^{ln}(S^{l}_{k}T^{i}_{n}
-S^{l}_{i}T^{k}_{n})\rbrace +{1\over 3}(C^{i}_{n}T^{n}_{j}T^{k}_{j}
-C^{k}_{n}T^{n}_{j}T^{i}_{j}). \eqno 35$$
The first term in (35) describes the transfer of angular momentum between 
neighbouring shells due to pressure
\footnote{
Note that  transfer of angular momentum due to self-gravity
is absent. Obviously, this is related to our simplified description
of the self-gravity forces in our model.}. The second term is obviously the tidal
torque term. The quantity in braces is equal to zero in the center 
and also at the boundary of the star. Therefore, if the tidal torque is absent,
the angular momentum is conserved. 

\section{Numerical work}

For our numerical work we would like to choose a simple polytropic model of the star
with the polytropic index $n=1.5$ and the specific heat ratio
$\gamma=5/3$. The star has the radius $R_*$ and mass $M_*$ equal to the
radius and mass of the Sun. It is assumed that the star is
moving along a marginally bound orbit 
in the field of the Kerr black hole. The same problem
has been discussed by Diener et al for the case of rather weak tidal interaction, 
and we use the results of this work for testing 
our model in the relativistic tidal field. We use the natural 'stellar' units
in our calculations and
representation of results: the dimensionless time
$\tau=\sqrt {GM\over R^{3}}$, the radius $\tilde R=r/R_{*}$, the mass coordinate
 $x=M/M_*$,
energy $\tilde E={E*R_{*}\over GM_*^{2}}$ and specific angular momentum 
$\tilde L=L/\sqrt{GM_*R_*}$. 

As was pointed out by IN, in the non-relativistic 
approximation the whole problem can be described by only two parameters:
the polytropic index $n$, and the quantity
$$\eta_{nr}= \sqrt{{M_*\over M_{bh}}{R_{p}^{3}\over R^{3}_{*}}}, \eqno 36$$ 
where $M_{bh}$ is the black hole mass and $R_{p}^{nr}$ is the pericentric separation
from the black hole calculated in the non-relativistic approximation. This 
quantity has been introduced by Press and Teukolsky in the linear theory of
tidal perturbations. The smaller this quantity is, the stronger the tidal 
encounter will be. For the relativistic field the situation is much more
complicated. For a fixed stellar model, 
the problem must be parametrised by the ratio of the star's
mass to the black hole mass in order to specify the importance of
relativistic corrections. The problem also depends on the dimensionless
rotational parameter $a$ of the black hole
\footnote{The dimensionless rotational parameter $a$ is determined by the
black hole angular momentum $J$ and its mass $M$: $a={cJ\over GM^{2}}$.}. 
The marginally bound star's orbit 
can be specified by its angular momentum $L_{orb}$ and the Carter integral
$Q$. In this paper we would like to consider the most interesting case
of  equatorial orbits, and therefore we set $Q=0$. We use the dimensionless
orbital angular momentum $\tilde L_{orb}=L_{orb}/ r_gc$, where
$r_{g}={2GM_{bh}\over c^{2}}$. 
Instead of using of
$\eta_{nr}$, Frolov et al (1994) proposed to use a more relevant
quantity
$$\eta=\eta_{nr}{(R_{p}/R_{p}^{nr})}^{3/2}, \eqno 37$$      
where $R_{p}$ is the minimal radial distance from the black hole calculated in
the relativistic approach. Note that the corresponding dimensionless
quantity $\tilde R_{p}=R_{p}/r_{g}$ can be expressed only in terms 
of $\tilde L_{orb}$ provided the rotational parameter $a$ is specified.    

For our calculations we use an explicit Lagrangian numerical scheme which is
similar to what was used by IN. 
The stability
criterion of our scheme is discussed in Appendix 1.
The essential differences of the new numerical
scheme from the scheme of IN are discussed in Appendix 2.
The results of our calculations are
presented in Fig 1-9. 

\begin{figure}
\vspace{8cm}
\hspace{-1cm}
\includegraphics{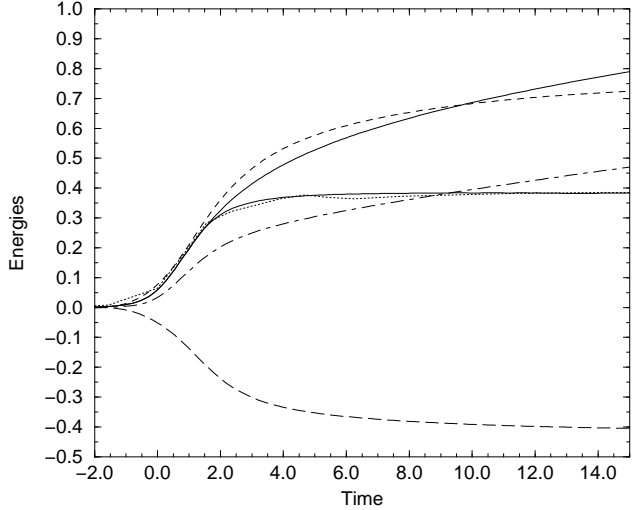}
\vspace{-1cm}
\caption{Evolution of total (solid curves), 
thermal (long dashed curve), gravitational (short dashed curve) and kinetic 
(dot-dashed curve) energies
with time. All energies are measured with respect to their equilibrium
values. The dotted curve represents the evolution of total energy of
the gravitationally bound part of the star in the model of Diener et al.}
\end{figure}  

In Figures $1-4$ we show the time dependence of different energies, 
the angular momentum, central density and mass lost by the
star. We choose $M_{bh}=1.0853\cdot 10^{7}M_{\odot}$, $a=0.9999$,
$\eta=1.6486$ and $\tilde L_{orb}=2.72945$. These parameters correspond
to the Diener et al model $5$ which has been intensely
investigated in 
3D finite difference simulations. 
Therefore, we use this calculation 
for testing our model and numerical scheme. In Fig. $1$ we show 
the time dependence 
of the total energy of the whole star (the upper solid curve) and
the total energy of gravitationally bound debris (the lower solid curve)
\footnote{The gravitationally bound debris is defined as the part of
the star where the sum of kinetic and gravitational
energies is less than zero.}.
The dot-dashed curve corresponds to the kinetic energy of the star, the 
dashed curve corresponds to the gravitational energy of the star and the
long-dashed curve
corresponds to the thermal energy of the star. In general, all curves look
very similar to the corresponding curves calculated by IN in the 
non-relativistic approximation. The dotted curve shows the time dependence 
of the total energy of the gravitational bound debris calculated
by Diener et al in the 3D calculations. One can see from this Figure 
that this curve almost coincides with our curve.

\begin{figure}
\vspace{8cm}
\hspace{-1cm}
\includegraphics{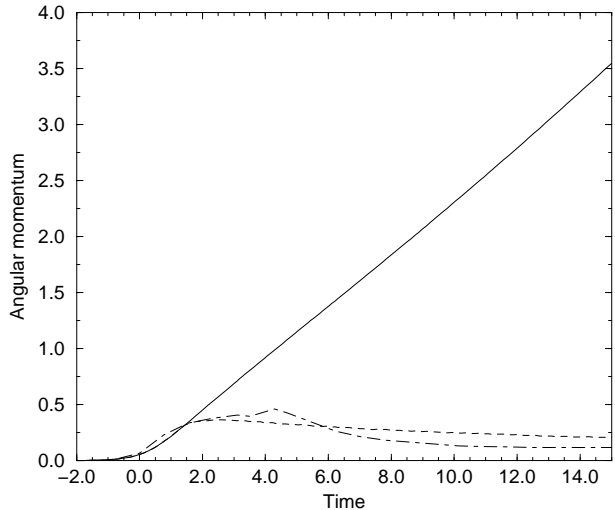}
\vspace{-1cm}
\caption{Evolution of the component of angular momentum perpendicular
to the equatorial plane of the black hole.}
\end{figure}    

In Fig. $2$ the total angular momentum
of the star (solid curve), the angular momentum of the gravitationally 
bound debris (dashed curve) and the same quantity calculated by Diener et
al (dot-dashed curve) are shown as a function of time. The total angular momentum
of the star grows monotonically with time, and is significantly larger than
the angular momentum of the gravitationally bound debris at the end of
the calculations. The angular momentum of gravitationally bound debris calculated
in the 3D simulations is close to our quantity for $\tau < 6$. 
Then, a sharp decrease of the angular momentum is observed and at the end
of the calculation the angular momentum of Diener et al is significantly
smaller than our quantity. The reason for this behaviour of the angular 
momentum found in the 3D simulations is not clear to us, and therefore
we cannot comment on this deviation.

\begin{figure}
\vspace{8cm}
\hspace{-1cm}
\includegraphics{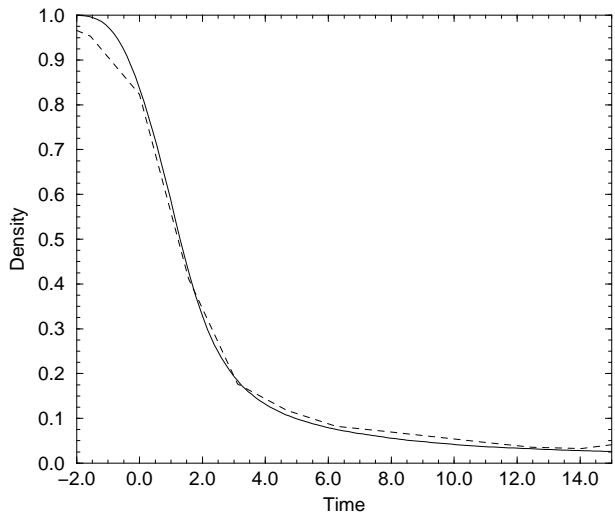}
\vspace{-1cm}
\caption{Evolution of the central density (expressed in units of the
central density of the unperturbed star) as a function of
time.}
\end{figure}
 
In Fig. $3$ the time evolution of the
the central density (expressed in units of the central density of 
the unperturbed star)  is shown. The solid curve corresponds to
our model and the dashed curve corresponds to the 3D simulations.
We can see again that these two curves are very similar. 

\begin{figure}
\vspace{8cm}
\hspace{-1cm}
\includegraphics{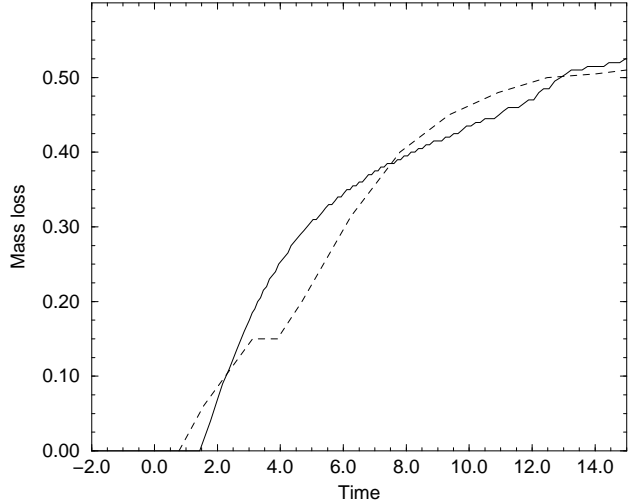}
\vspace{-1cm}
\caption{The amount of mass (expressed in units of 
the stellar mass) lost by the star 
during the tidal encounter as a function of
time.}
\end{figure} 

In Fig. $4$
we show the amount of mass lost by the star (expressed in units of 
the stellar mass) with time.  The solid curve and the dashed curve
correspond to our model and the 3D simulations, respectively.
The asymptotic value of the mass loss is almost the same in  
both cases and is about 0.5. However, we would like to note that
there is a significant ambiguity in determining the gravitationally
bound gas in the 3D simulation (Diener et al). If one considers all the gas
elements which left the computational domain with velocities less than the escape
velocity as still being present in the debris, the mass loss would be
significantly less with an asymptotic value $\sim 0.3$. In general, we 
think that our model shows very good similarities to the 3D simulations
for the parameters chosen in this computation.

\begin{figure*}
\vspace{10cm}
\centerline{
\hspace{-8cm}
\includegraphics{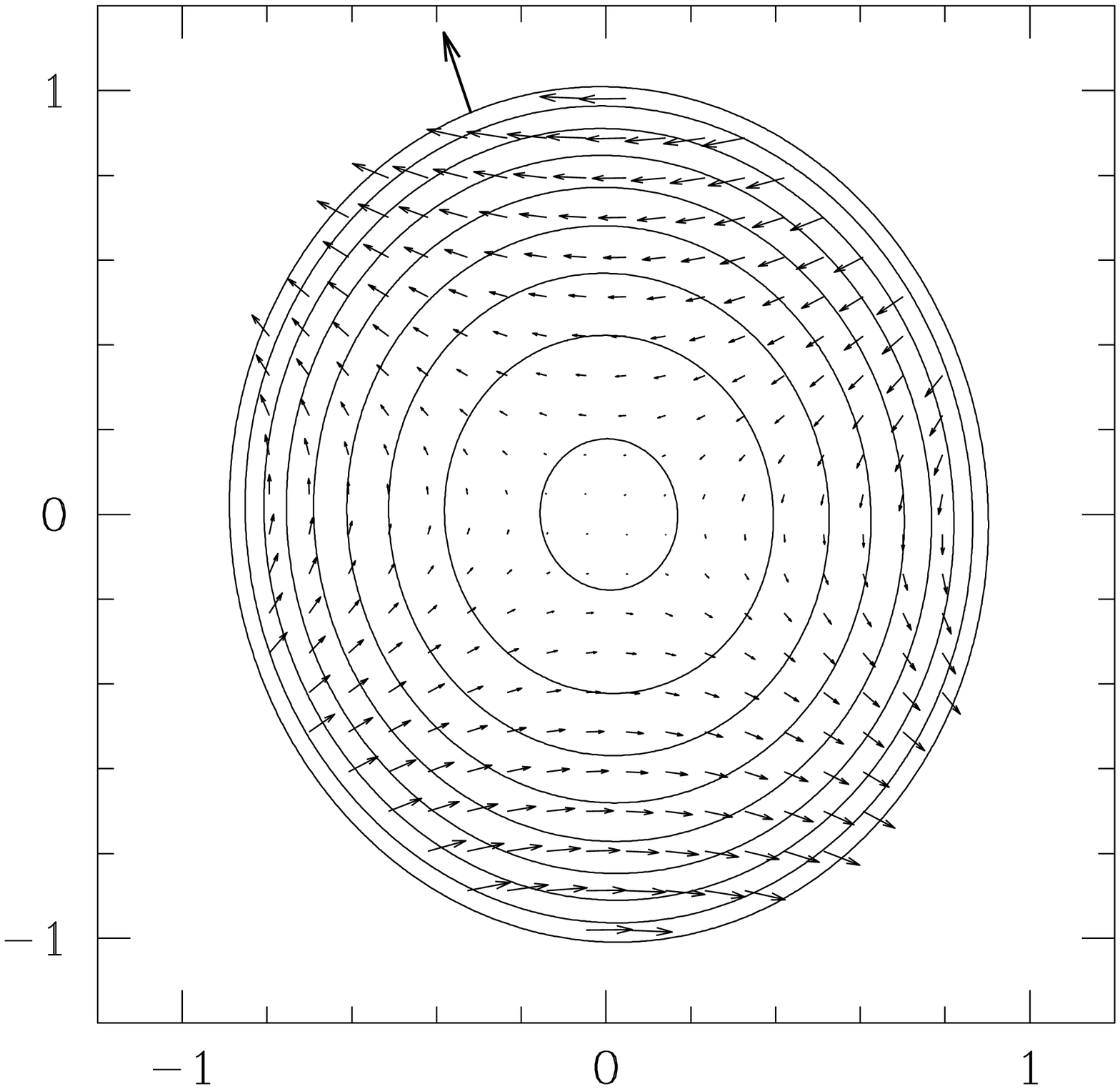}
\hspace{8cm}
\includegraphics{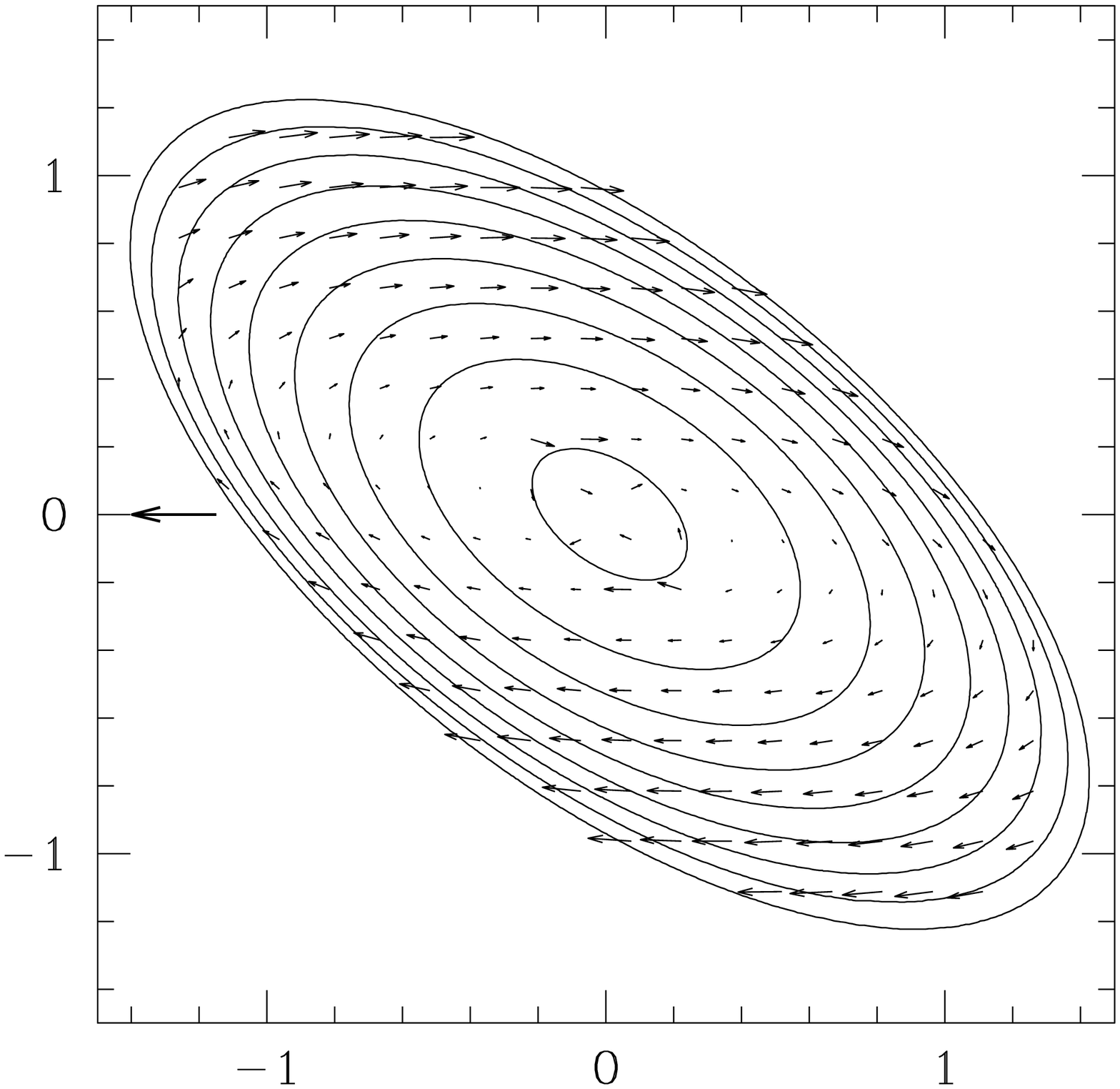}}
\vspace{7cm}
\centerline{
\hspace{-8cm}
\includegraphics{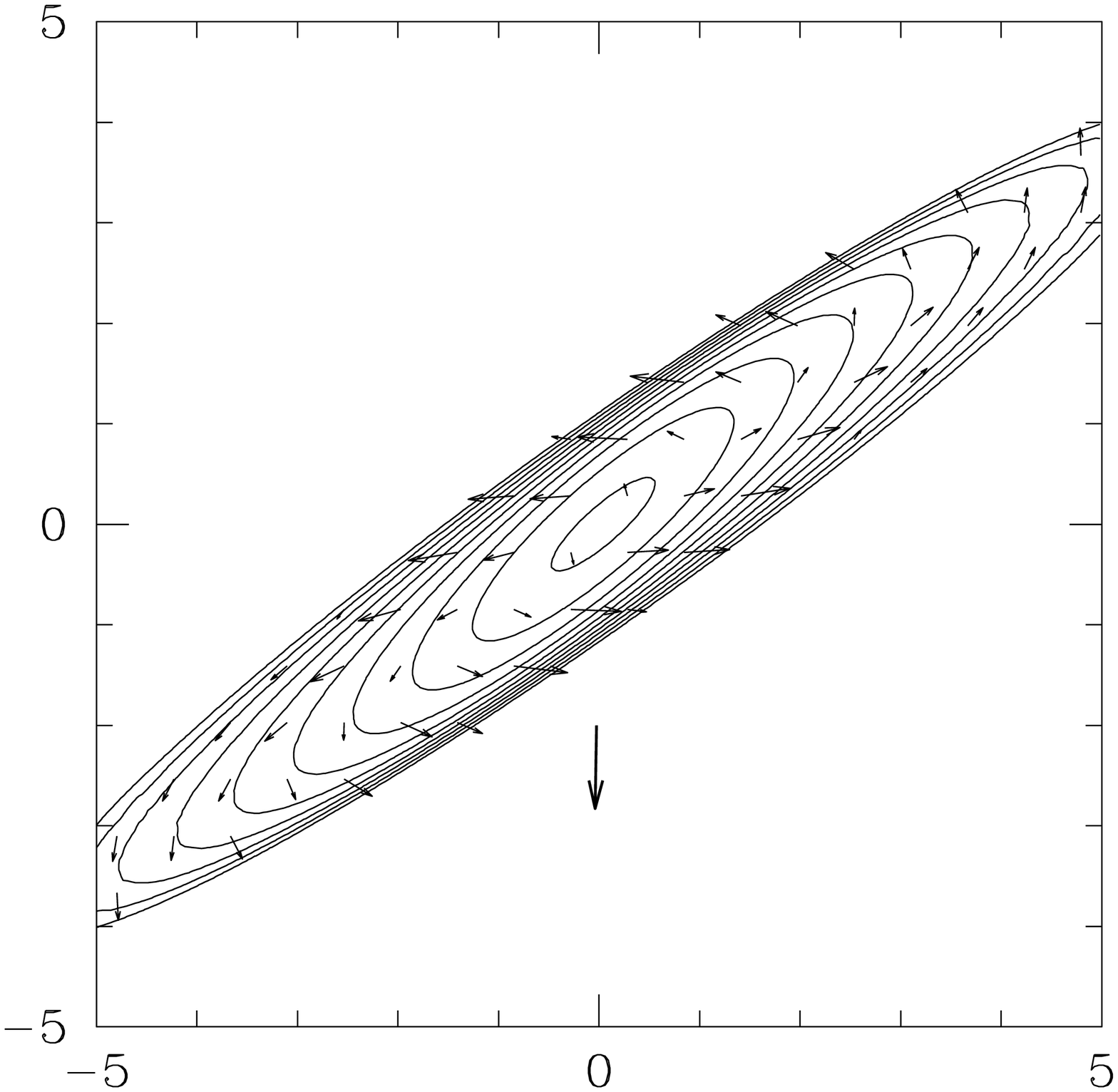}
\hspace{8cm}
\includegraphics{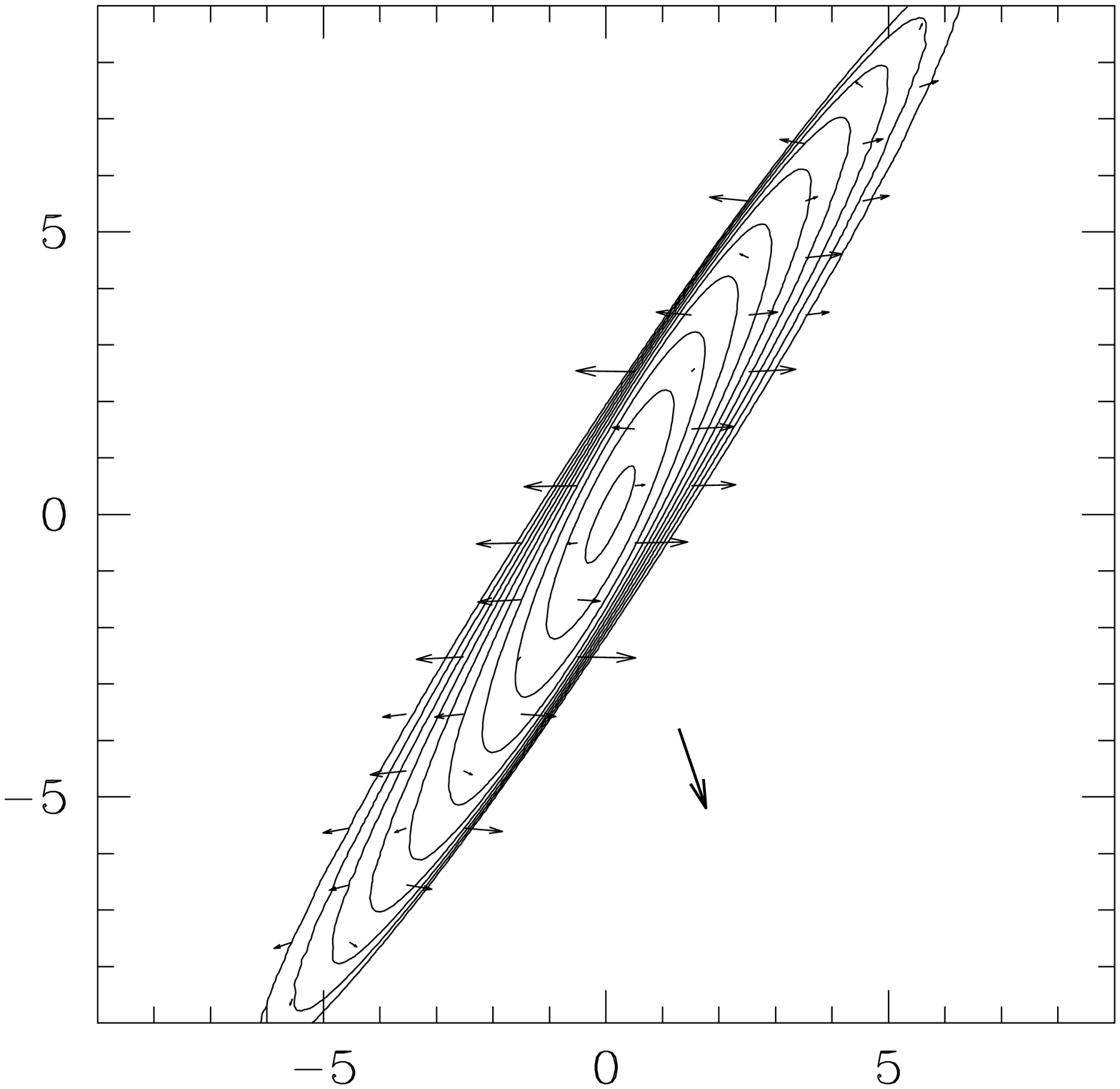}}
\vspace{-1.5 cm}
\caption{ The time evolution of the velocity field and
density contours in the orbital X-Y plane for the $\eta=1.6486$ encounter.
The spatial scales are expressed in units of the unperturbed stellar
 radius $R_*$. The large arrow points toward the black hole.
The innermost contour corresponds to 95\% of the central density.
 At each subsequent contour  density decreases by a 
factor of 1.5 . {\it Upper left-hand panel:}
 $\tau=-2$ ($R/R_p=1.5$),{\it Upper right-hand panel:}
$\tau=0$ ($R/R_p=1$),{\it Bottom left-hand panel:} $\tau=3$ ($R/R_p=1.9$),
{\it Bottom right-hand panel:} $\tau=5$ ($R/R_p=2.73$). 
The small arrows show the direction and the 
relative magnitudes of the velocities.}
\end{figure*}
           
Figure $5$ shows the time evolution of the velocity field and
density contours in the orbital (XOY) plane 
for the same model. Axis (OX) is directed opposite to the black hole during
the pericenter passage. Time $\tau=0$ corresponds to the time of
 the pericenter passage.
In this figure one can see that the density contours lose 
their spherical shape with time and  elongate. The size of the outermost 
 contour at $\tau=5$ is about 10 times
larger than at $\tau=-2$. In  the beginning,  the
contours expand in the direction of the black
hole, but as the star approaches the pericenter of its orbit,
 they start to lag behind.  The lag of the innermost contours is slightly
less than that  of the outer ones.  
The distribution of the velocities in the star 
is represented by arrows whose lengths are proportional 
to the velocity magnitude.

Now let us discuss some simple properties of the parameter space of the problem.
Figures (6-9) show dependencies of the mass lost by the star, the total
energy contained in the gravitationally bound debris and the angular momentum
of the debris on the orbital angular momentum $\tilde L_{orb}$
for two values of the rotational parameter of the black hole: $a=0.9999$ and 
$a=0$.
The mass of the black hole and other orbital parameters 
are the same as in the previous calculations. 
The largest orbital angular momentum corresponds to $\eta=2$, the smallest angular momentum
corresponds to the total disruption of the star (we denote the respective value of
$\eta$ as the 'critical' $\eta_{cr}$).  
\begin{figure}
\vspace{7cm}
%\hspace{-1cm}
\includegraphics{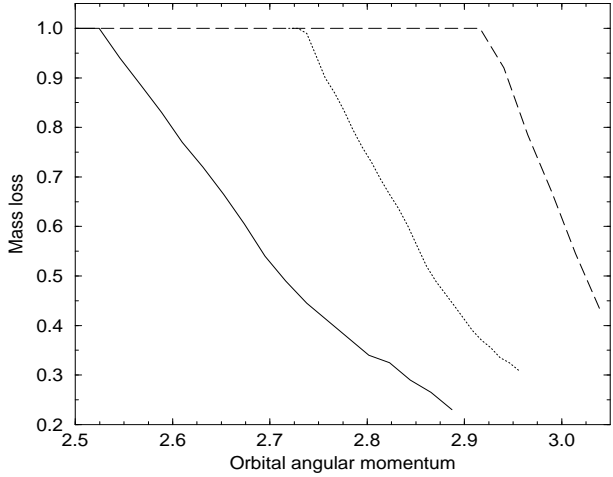}
\vspace{-0.5cm}
\caption{
The mass lost by the star after a fly-by of the black hole
is shown as a function of the absolute magnitude of the dimensionless orbital momentum. 
The case  $M_{bh}=1.0853\cdot 10^{7}M_{\odot}$ is considered.
The solid and dashed curves correspond to 
$a=0.9999$, positive and negative orbital angular momenta, respectively.
The dotted curve correspond to $a=0$.}
\end{figure}

\begin{figure}
\vspace{8.5cm}\includegraphics{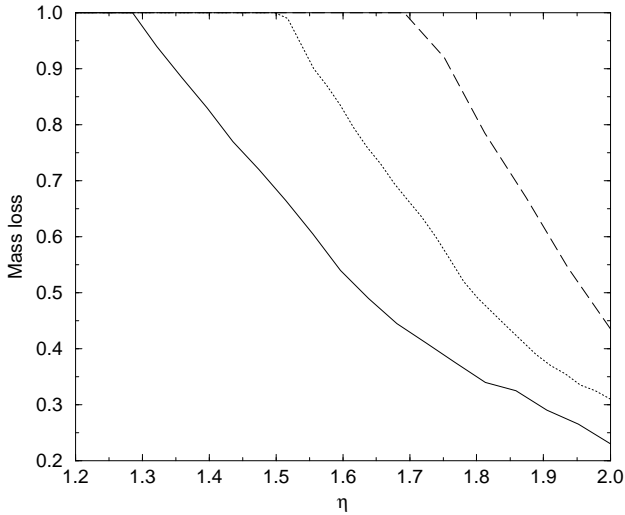}
\vspace{-0.5cm}
\caption{
Same as Figure 6, but the mass lost by the star is shown as a function of the
parameter $\eta$.}
\end{figure}

Figures (6,7) show the dependence of the mass lost by the star on the absolute magnitude
of the orbital angular 
momentum and parameter $\eta$, respectively. The solid and dashed curve correspond to positive and negative angular momenta
and are calculated for $a=0.9999$.
The dotted curve is calculated for $a=0$.  
It is seen from the figures that
the stars with negative orbital angular momenta are disrupted much more effectively than the
stars with positive angular momenta 
\footnote{ 
Note that this effect has been discussed by Beloborodov et al (1992) in the framework of a rather
naive criterion for tidal disruption.}. It is obvious that similar curves calculated for
smaller rotational parameters of the black hole (but with the same mass) must lie between
the solid and dashed curves.    
We have $\eta_{cr}=1.5$ for $a=0$ and $\eta_{cr}=1.28,\quad 1.69$ for $a=0.9999$ and positive
and negative orbital angular momenta, respectively.

\begin{figure}
\vspace{7cm}
%\hspace{-1cm}
\includegraphics{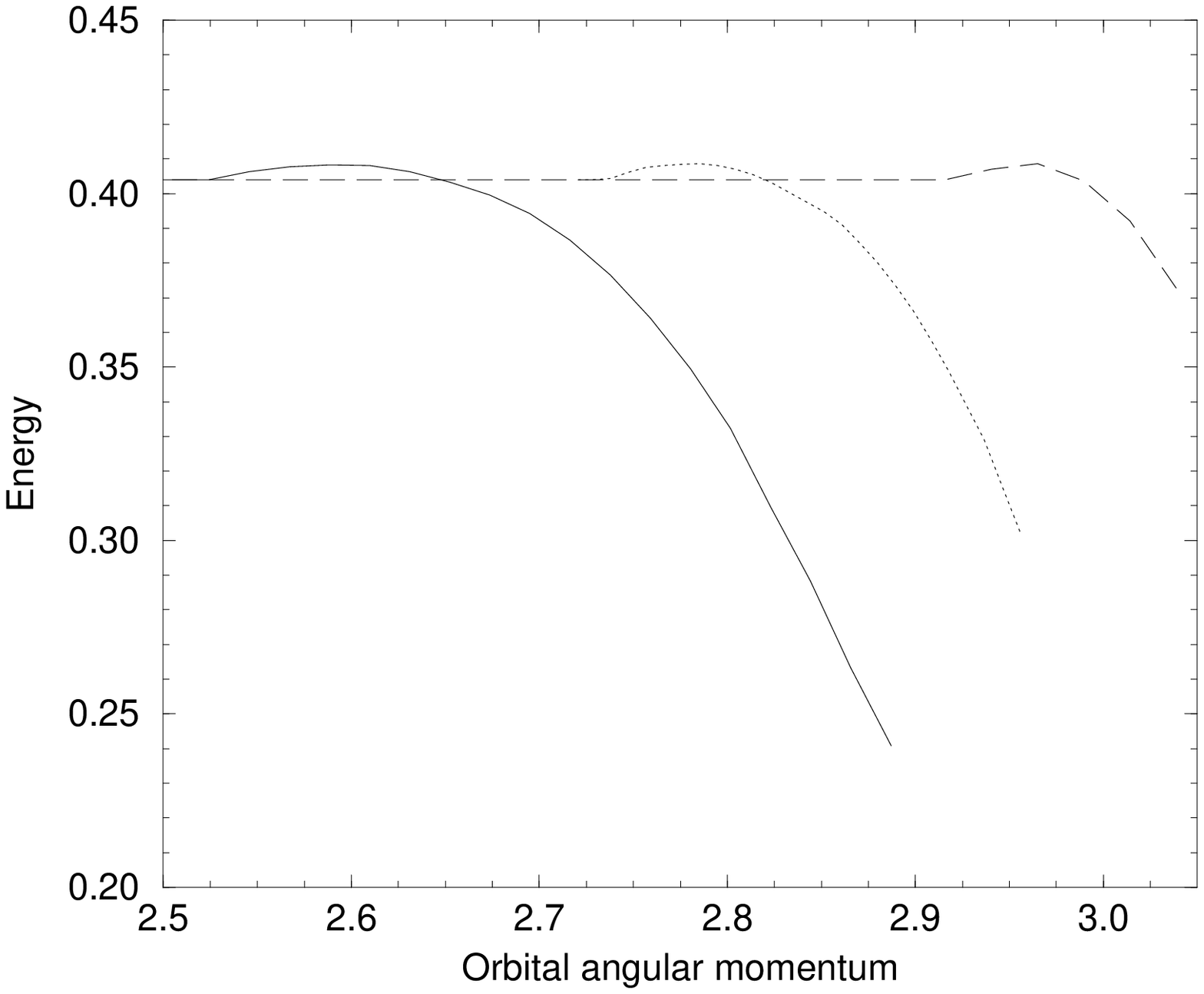}
\vspace{-0.5cm}
\caption{
The energy of gravitationally bound debris is shown as a 
function of the absolute magnitude of the dimensionless orbital angular momentum. 
The solid and dashed curves correspond to 
$a=0.9999$, positive and negative orbital angular momenta, respectively.
The dotted curve correspond to $a=0$.   
}
\end{figure}

\begin{figure}
\vspace{7cm}
%\hspace{-1cm}
\includegraphics{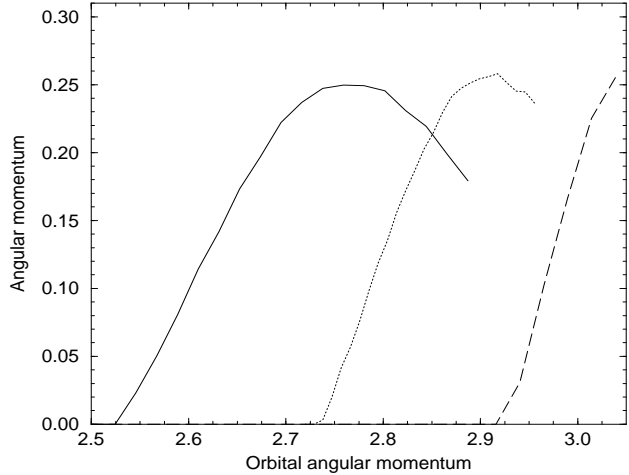}
\vspace{-0.5cm}
\caption{
The angular momentum of gravitationally bound debris is shown as a 
function of the absolute magnitude  of the dimensionless orbital angular momentum. 
The solid and dashed curves correspond to 
$a=0.9999$, positive and negative orbital angular momenta, respectively.
The dotted curve correspond to $a=0$.
}
\end{figure} 

In Figure 8 we show the dependence of the total energy of the gravitationally bound
debris calculated after the fly-by of the black hole on the value of the dimensionless
orbital angular momentum. Similar to Figures 6,7 representing the mass loss, Figure 8 shows
that the stars moving on orbits with negative orbital angular momentum are perturbed more 
effectively than the stars moving on orbits with positive angular momentum. It is 
interesting to note that the asymptotic value of the total energy in the limit of
small orbital angular momentum (corresponding to full disruption of the star) is
a nonzero quantity which depend neither on the spin of the black hole nor
on the sign of the orbital angular momentum (see Figure 8). This could be explained as
follows: after the fly-by the stellar gas leaves the star with 
almost zero specific total energy (i. e. parabolic velocities). So, the out-flowing gas 
does almost not carry specific energy, and the energy of the gravitationally bound part of the star
is conserved. In Figure 9 we show the component of the angular momentum of the star
perpendicular to the orbital plane after the fly-by as a function of
the dimensionless orbital angular momentum. Contrary to the total energy, the angular momentum 
of the gravitationally bound debris is a non-monotonic 
function of the dimensionless orbital angular momentum.

\section{Discussion}

In this paper we construct a self-consistent variant of the new model of a tidally perturbed
or tidally disrupted star proposed by Ivanov and Novikov 2001. 
The model allows researchers to calculate the outcome
of the tidal disruption event with the help of a one-dimensional Lagrangian numerical scheme. Therefore,
it is much faster than the conventional numerical 3D approach, and it could be evolved for much 
longer time. We use the model in numerical calculations of the tidal interaction of an $n=1.5$ polytropic
star with a Kerr black hole of mass $10^{7}M_{\odot}$. We compare the results of our calculations 
with the results of finite difference $3D$ calculations of the same problem and find a very
good agreement between them. Then, we consider dependencies of the main characteristics of the tidally 
perturbed star after a fly-by of the black hole in the equatorial plane on the value of the orbital
angular momentum. We find that the stars with negative orbital angular momentum are perturbed more
effectively than the stars with positive orbital angular momentum. We also briefly discuss the dependence of
the outcome of the tidal encounter on the spin of the black hole.

As it was demonstrated in the present work (see also IN), the model gives results which in certain
cases coincide almost completely with results of $3D$ calculations. On the other hand, the dynamical
equations of our model cannot be reduced to the exact hydrodynamical equations. Therefore, a natural
question arises: why is the agreement between the two approaches so good? The possible explanation might be
as follows. The key assumption of our model consists in using elliptical shells for the description of 
the shape of the star evolving under the influence of the tidal field. It seems that 
the quadrupole dependence of the field of tidal forces on the angular coordinates and the 
special algebraic properties of the tidal tensor could justify such assumptions at least for large
scale hydrodynamical motions induced in the star. This could answer qualitatively the question why
different elements of the star which are not in causal contact evolve in such a way that the
elliptical form of the shells is always maintained.  
 
Now let us discuss the problems of our model. 
At first  we discovered in our numerical calculations our numerical is scheme is slowly 
unstable for stellar models with a more sharp density contrast. For example, in the case of
an $n=3$ polytrope it takes several characteristic 'stellar' times for the instability to halt the
computations. Since our numerical scheme has been written in a rather naive manner, we expect
that a more sophisticated numerical scheme (e.g. an implicit scheme) could resolve this
difficulty. There is a more fundamental problem of our model. We managed to obtain the 
distribution of pressure and density across the star in a simple form only for a polytropic 
star. Therefore, it is not clear for us how to use our model for the case of a more realistic
stellar gas
\footnote{  
Note that the variant of the model considered by IN is free from this difficulty.}.

In the present paper we do not make an attempt to comprehensively survey the 
parameter space of the
relativistic tidal problem and to calculate cross-sections of 
different kind. This will be treated in future work.  

\section*{Acknowledgements}
We are grateful to J. C. B. Papaloizou and A. G. Polnarev for very useful 
discussions, and Martin Goetz for very helpful comments. 
This work has been supported in part by RFBR grant 00-02-16135, 
in part by the Danish Research Foundation through its 
establishment of the Theoretical Astrophysics Center, and in part 
by the Danish Natural Science Research Council through grant N 9701841.

\appendix

\section{Propagation of small perturbations and the time step constraint}

In a standard approach the time step constraint 
must follow from a stability analysis of
numerical schemes. However,
the standard stability analysis of our numerical scheme is rather complicated and
therefore we do not use it in our paper. Following IN, we constrain our time
step by the condition 
$$\delta t = {\alpha \delta M\over c_{smax}}. \eqno A1$$
Here $\alpha < 1$ is a parameter. $c_{smax}$ is greater than or equal to the velocity of
propagation of a small perturbation (with respect to the mass coordinate) $c_{s}$
calculated in analytical linear approximation: $c_{smax} \ge c_{s}$.
For the stellar gas we assume
the equation of state of an ideal gas with polytropic index $\gamma$.

To estimate $c_{smax}$ we decompose our dynamical variable ${\bf T}$ in a 
background part and a perturbation: ${\bf T}= {\bf T}_{0}+
{\bf t}$, 
where the perturbation ${\bf t}$ is assumed to be of standard oscillatory
form:
$${\bf t}= \tilde {\bf t}e^{i(\omega t+kM)}. \eqno A2 $$
For the velocity  $c_{s}$ we have
$$c_{s}={\omega \over k}$$

The dynamical equation for the perturbation follows directly from equations (21)
and (17). Neglecting the dependence of the background quantities on the mass coordinate
and taking into account only the second term on the right hand side of equation (21),
we have:
$${\partial^{2}\over \partial t^{2}} t^{i}_{j}=24\pi^{2}\gamma g^{(1-\gamma)}r_{0}^{2(\gamma +1)}p_{0}\rho_{0}S^{l}_{i}
H^{ljkn}(S^{k}_{m}{\partial^{2} \over \partial M^{2}}t^{m}_{n}
+S^{n}_{m}{\partial^{2} \over \partial M^{2}}t^{m}_{k}), \eqno A3$$
with the symmetric tensor
$$H^{ljkn}=\int {d\Omega \over 4\pi}{e_{0}^{l}e_{0}^{j}e_{0}^{k}e_{0}^{n}
\over {(e_{0}^{s}e_{0}^{\rho}R^{\rho}_{s})}^{(\gamma +1)}}. \eqno A4$$ 
Substituting (A2) in (A4), one can obtain a set of algebraic equations. Then, the
usual compatibility condition gives the value of the velocity $c_{s}$.
However, this is too complicated for our purposes. To estimate the upper limit
of the velocity, we can use an upper limit estimate for the tensor $H^{ljkn}$:
$$H^{ljkn} \le {1\over 15f_{min}}(\delta^{lj}\delta^{kn}+\delta^{lk}\delta^{ln}+\delta^{ln}\delta^{lk}), \eqno
A5$$ 
where $f_{min}$ is the minimal eigenvalue of the matrix $\bf R$. 
Substituting (A5) in (A3), and using (A2), we have:
$$a^{l}_{j}={A\over c_{s}^{2}}S^{l}_{i}S^{k}_{i}(a^{j}_{k}+a^{k}_{j}+\delta^{k}_{j}a^{n}_{n}) \eqno A6$$
where
$$a^{l}_{j}=S^{l}_{i}\tilde t^{i}_{j}, \eqno A7$$
and
$$A={{(4\pi )}^{2}\over 5}\gamma g^{(1-\gamma )}{r_{0}^{2(\gamma +1)}\over f_{min}^{\gamma +1}}p_{0}\rho_{0}.
\eqno A8$$
Let us  assume the matrix ${\bf a}$ to be symmetric: ${\bf a}={\bf a}^{T}$. 
In this case both matrices $a^{l}_{j}$ and $S^{l}_{i}S^{j}_{i}$ can be diagonalized by the same 
transformation. Taking into account that the quantities $a^{-2}_{i}$ are the eigenvalues of the 
matrix $S^{l}_{i}S^{j}_{i}$, equation (A6) can be reduced to the form:
$$c_{i}={A\over c_{s}^{2}}a^{-2}_{i}(2c_{i}+\sum^{i=3}_{i=1}c_{i}), \eqno A9$$
where the quantities $c_{i}$ are the eigenvalues of the matrix ${\bf a}$.    
Equation (A9) has nontrivial solutions if and only if
$$1=\sum^{i=3}_{i=1}{1\over (a_{i}^{2}x-2)}, \eqno A10$$
where $x=c_{s}^{2}/A$. Obviously, equation (A10) 
gives an implicit dispersion relation.
A simple analysis of equation (A10) shows that all roots of (A10)
are larger than
$$f=3({1\over a^{2}_{1}}+{1\over a^{2}_{2}}+{1\over a^{2}_{3}}), \eqno A11$$  
and we have
$$c_{s} > \sqrt {fA}. \eqno A12$$
Therefore, we use the condition
$$\delta t = {\alpha \delta M \over \sqrt {fA}} \eqno A13$$
in order to constrain our time step. 
  
\section{The numerical scheme}

The numerical scheme used in the computations is very similar to what was used by IN
and we address the reader to Appendix B of IN for a comprehensive description of the scheme.
The main difference is determined by the fact that we now calculate the exact distributions
of pressure and density across the star contrary to the approximate treatment of these 
quantities by IN. These quantities
determine the thermal terms (16,17) which in turn determine the action of 
pressure forces in our model. 
In this Appendix we outline the numerical evaluation of the thermal terms (16,17).

The calculation of these terms can be subdivided in two steps. First, with help of a special
subroutine we find the eigenvalues and eigenvectors of the shear matrix ${\bf R}$
(i. e. the quantities $f_m$ and the matrix ${\bf O}$, see eq. (6)).
The pressure tensor (17) is diagonal in the
frame of eigenvectors of the matrix ${\bf R}$. 
Therefore, we have to evaluate numerically only four quantities: 
three eigenvalues of the tensor (17) and the quantity (16).
It turns out
that in the case of polytropic stellar gas, the eigenvalues of the pressure tensor
are proportional to the following integrals:
$$I_i\equiv \int^{2\pi}_{0}d\phi \int^{\pi}_{0}d\theta sin( \theta){ 
e^{2}_{i} \over 
{(f_1cos^{2}(\theta)+sin^{2}(\theta )(f_2cos^{2}(\theta)+f_3sin^{2}(\theta)))}^{\gamma}},
\eqno B1$$
where $e_{i}$ are the direction cosines and $\gamma=5/3$ is the specific heat ratio.
Quantity (16) is proportional to
$$I_0\equiv \int^{2\pi}_{0}d\phi \int^{\pi}_{0}d\theta sin( \theta){ 
1 \over 
{(f_1cos^{2}(\theta)+sin^{2}(\theta )(f_2cos^{2}(\theta)+f_3sin^{2}(\theta)))}^{\gamma-1}},
\eqno B2$$  
Integrals (B1) and (B2) are evaluated numerically by a separate program and 
tabulated as functions of 
the ratios $f(2)/f(1)$ and $f(3)/f(1)$. For small and large values of the ratios
we use analytical representations of these integrals in terms of series.

\end{document}